\documentclass[useAMS,usenatbib, doublecolumn]{mn2e}
\usepackage{epsfig}
\usepackage{amsmath}
\usepackage{amssymb}

\title[Evolution of BHIMXBs]
{Evolution of black-hole intermediate-mass X-ray binaries: the
influence of a circumbinary disc}
\author[Wen-Cong Chen and Xiang-Dong Li]{Wen-Cong Chen\thanks{E-mail:
chenwc@nju.edu.cn} and Xiang-Dong
Li\thanks{E-mail:lixd@nju.edu.cn }\\
Department of Astronomy, Nanjing University, Nanjing 210093, China
}
\begin{document}

\date{Accepted . Received ; in original form }

\pagerange{\pageref{firstpage}--\pageref{lastpage}} \pubyear{2006}

\maketitle

\label{firstpage}

\begin{abstract}
\citet{just06} recently suggested that black-hole low-mass X-ray
binaries (BHLMXBs) with short orbital periods may have evolved
from black-hole intermediate-mass X-ray binaries (BHIMXBs). In
their model the secondaries in BHIMXBs are assumed to  possess
anomalously high magnetic fields, so that magnetic braking can
lead to substantial loss of angular momentum. In this paper we
propose an alternative mechanism for orbital angular momentum loss
in BHIMXBs. We assume that a small fraction $\delta$ of the
transferred mass from the donor star form a circumbinary disc
surrounding the binary system. The tidal torques exerted by the
disc can effectively drain orbital angular momentum from the
binary. We have numerically calculated the evolutionary sequences
of BHIMXBs, to examine the influence of the circumbinary disc on
the binary evolution. Our results indicate when $\delta\la
0.01-0.1$ (depending on the initial orbital periods), the
circumbinary disc can cause secular orbital shrinking, leading to
the formation of compact
 BHLMXBs, otherwise the orbits always expand during the evolution.
This scenario also suggests the possible existence of luminous,
persistent BHLMXBs, but it suffers the same problem as in
\citet{just06} that, the predicted effective temperatures  of the
donor stars are significantly higher than  those of the observed
donor stars in BHLMXBs.
\end{abstract}


\section{Introduction}
If the gravitational mass of a compact star exceeds the maximum
value $\sim 2-3M_{\odot}$ of a neutron star \citep[e.g.][]{rhoa74},
this object should be taken as a black hole (BH) candidate. There
exist currently around twenty stellar-mass BH candidates
\citep[see][]{lee02,oros02,casa06}, all of them are located in
binary systems where their dynamical masses can be availably
estimated. Nine of these systems are defined as compact BH X-ray
binaries (BHXBs) with short orbital periods ($\la$ 0.5 d) and
low-mass donors ($< 1M_{\odot}$) \citep{lee02,ritt03,pods03}. It was
estimated that there may exist $\ga1000$ compact BHXBs in the Galaxy
\citep{wije96,roma98}.

The short orbital periods of BH low-mass X-ray binaries (BHLMXBs)
imply that they must have undergone secular orbital angular momentum
loss. If their progenitor systems contains a low-mass secondary
initially, it is not clear whether the secondary star has enough
energy to eject the envelope of the black hole progenitor during the
common envelope evolution phase \citep{pods03,just06}. This
difficulty can not be dealt with by assuming a intermediate-mass
secondary, which have radiative envelope and are not expected to be
subject to magnetic braking \citep{kawa88}, since the binary orbits
will be widen when mass is transferred from the less massive
secondary to the more massive BH \citep[see][]{huan63}.
\citet{eggl86} suggested that these systems may be evolved from
triple systems via the merging of the two massive components.
\citet{pods95} proposed that massive Thorne-\.{Z}ytkow objects could
lead to the birth of BHs with a low-mass donor star. \citet{just06}
summarized the previous proposals for the formation of BHLMXBs. They
instead suggested that BHXBs containing intermediate-mass Ap and Bp
donor stars, which possess strong magnetic fields, may be driven to
compact BHLMXBs via magnetic braking with an irradiation-driven wind
from the donor star \citep[see however,][]{yun06}.

Are there additional angular momentum loss mechanisms for BHXBs
besides magnetic braking? Here we explore an alternative
possibility. It is well known that some fraction of the transferred
matter from the donor star may leave the system in various ways
during the mass exchange process  \citep{heu94}. In particular, for
cataclysmic variables (CVs), \citet{spru01} and \citet{taam01}
suggested that part of the outflow may be in the form of a
circumbinary (CB) disc, and investigated the influence of the CB
disc on the evolution of CVs. The similar idea was adopted by
\citet{mey01} for BHXBs to regulate the mass transfer rates, in
which the CB disc could result from the remnants of the previous
common-envelope phase. In our previous works we have investigated
the evolution of neutron star (NS)  LMXBs and Algol type binaries
with a CB disc \citep{chen06a,chen06b}. These works indicate that
the presence of a CB disc can accelerate the evolution process of
the binary systems, enhance the mass transfer rates, and lead to
secular orbit shrinking under certain conditions.

The aim of this work is to study the influence of the CB disc on
the evolution of BH intermediate-mass X-ray binaries (BHIMXBs),
and explore the possibility of BHIMXBs being the progenitor
systems of short period BHLMXBs. The structure of this paper is as
follows. In section 2 we describe the adopted orbital angular
momentum loss mechanisms in the evolution model of BHIMXBs. In
section 3 we present the numerically calculated results for the
evolutionary sequences. We summarize and discuss the uncertainties
in our model in section 4.

\section[]{Description of the Model}

Although some BHXBs may form dynamically in globular clusters
\citep{rapp01}, we only consider those evolved from the primordial
massive binary systems. We focus on the evolution of BHXBs
consisting of an intermediate-mass donor star (of mass $M_{\rm
d}\sim 3-5 M_{\odot}$) and a black hole (of mass $M_{\rm BH}$).
Neglecting the spin angular momenta of both components, we
consider three types of angular momentum loss from the binary
system (in all calculations, magnetic braking mechanism is not
included), which are described as follows.

\subsection[]{Gravitational wave radiation}
For a compact BHXBs  gravitational wave radiation is able to carry
away orbital angular momentum effectively and lead to mass
transfer. The angular momentum loss rate due to gravitational
radiation is given by \citep{lan62}:
\begin{equation}
\dot{J}_{\rm GR}=-\frac{32}{5}\frac{G^{7/2}}{c^{5}}\frac{M_{\rm
BH}^{2}M_{\rm d}^{2}M^{1/2}}{a^{7/2}},
\end{equation}
where $G$ is the gravitational constant, $c$ the speed of light,
$M=M_{\rm BH}+M_{\rm d}$ the total mass of the binary, $a$ the
binary separation given by the Kepler's third law
$a=(GM/\Omega^{2})^{1/3}$, where $\Omega$ is the orbital angular
velocity of the binary system.

\subsection[]{Isotropic winds}
Mass and orbital angular momentum loss may occur during rapid mass
transfer phase, since the mass accretion rate of a black hole is
limited by Eddington mass-accretion rate\footnote{We do not
consider mass and angular momentum loss due to the wind mass loss
from the donor star, which is not important for intermediate-mass
stars. Irradiation effect on the stellar winds is also not
included.}. For spherical accretion, this maximum accretion rate
can be derived from the equation that gravity can balance the
radiation pressure:
\begin{equation}
  \dot{M}_{\rm Edd}=\frac{4\pi GM_{\rm BH}}{\kappa c\eta_{\rm E}},
\end{equation}
where $\kappa=0.2(1+X)\rm cm^{2}g^{-1}$ is the electron scattering
opacity for a composition with hydrogen mass fraction X
\citep{kipp90}. The energy release efficiency $\eta_{\rm E}$ of
disc accretion onto black hole can be approximately written as:
\begin{equation}
\eta_{\rm E}=1-\sqrt{1-\left(\frac{M_{\rm BH}}{3M^{0}_{\rm
BH}}\right)^{2}}
\end{equation}
as $M_{\rm BH}<\sqrt{6}M^{0}_{\rm BH}$, where $M^{0}_{\rm BH}$ is
the initial mass of the BH  \citep[see, e.g.][]{bard70,king99}. From
the above equations, \citet{pods03} obtained the expression of the
Eddington mass-accretion rate for a disc-fed BH:
\begin{equation}
  \dot{M}_{\rm Edd}\simeq 2.6\times 10^{-7} M_{\odot} {\rm yr^{-1}}
  \left(\frac{M_{\rm BH}}{10M_{\odot}}\right)
  \left(\frac{0.1}{\eta_{\rm E}}\right)\left(\frac{1.7}{1+X}\right).
\end{equation}

In our calculations we assume that the transferred matter in
excess of the Eddington accretion rate is ejected in the vicinity
of the BH in the form of isotropic winds, carrying away the
specific orbital angular momentum of the BH. We introduce the
parameter $f$ to describe the fraction of mass loss from the
binary system, defined by the following relations:
\begin{equation}
\dot{M}=\dot{M}_{\rm d}f,
\end{equation}
and
\begin{equation}
\dot{M}_{\rm BH}=\dot{M}_{\rm d}(f-1),
\end{equation}
where $\dot{M}$ is the total mass-loss rate from the system,
$\dot{M}_{\rm d}$ is the mass transfer rate from the secondary,
and $\dot{M}_{\rm BH}$ is the mass accretion rate of the black
hole, respectively. If the mass transfer rate $|\dot{M}_{\rm
d}|\leq \dot{M}_{\rm Edd}$, $f=0$, else $f=1+\dot{M}_{\rm
Edd}/\dot{M}_{\rm d}$.

Based on the above equations and assumptions, the angular momentum
loss rate via isotropic winds is
\begin{equation}
\dot{J}_{\rm IW}=f\frac{\dot{M}_{\rm d}M_{\rm d}}{MM_{\rm BH}}J,
\end{equation}
where
\begin{equation}
J=a^{2}\mu\Omega
\end{equation}
is the total orbital angular momentum and $\mu=M_{\rm d}M_{\rm
BH}/{M}$ is the reduced mass of the binary system, respectively.

\subsection[]{CB disc}

We further assume that a small fraction $\delta$ of the material
overflowed from the donor star's Roche-lobe {\em always} feeds into
the CB disc rather accretes onto the BH. Then $\dot{M}_{\rm d}$ in
Eqs.~(5)-(7) should be replaced by $\dot{M}_{\rm d}(1-\delta)$, the
BH mass accretion rate is $\dot{M}_{\rm
BH}=-(1-f)(1-\delta)\dot{M}_{\rm d}$, and the total mass loss rate
$\dot{M}=\dot{M}_{\rm d}(f+\delta-f\delta)$. Tidal torques are then
exerted on the CB disc extracting orbital angular momentum from the
binary system \citep[see][]{taam01}.

Similar as in \citet{taam01}, \citet{spru01}, and \citet{chen06b},
the viscous torque exerted at the inner edge $r_{\rm i}$ of the CB
disc\footnote{In this subsection we use the subscript i to denote
quantities evaluated at the inner edge $r_{\rm i}$ of the CB
disc.} on the BHXB can be shown to be :
\begin{equation}
T_{\rm i}\equiv\dot{J}_{\rm CB}=\gamma\Omega
a^{2}\delta\dot{M}_{\rm d}\left(\frac{t}{t_{\rm vi}}\right)^{1/3},
\end{equation}
where $\gamma^{2}=r_{\rm i}/a$, $t$ is the time since the onset of
Roche lobe overflow (RLOF). In the standard $\alpha$ viscosity
prescription \citep{Shak73}, the viscous timescale $t_{\rm vi}$ at
the inner edge in the CB disc is given by:
\begin{equation}
t_{\rm vi}=\frac{4\gamma^{3}}{3\alpha\Omega\beta^{2}},
\end{equation}
where $\beta=H_{\rm i}/r_{\rm i}$, $\alpha$ and $H_{\rm i}$ are the
viscosity parameter and the scale height of disc, respectively.

Combine Eqs.~(9) and (10), the angular momentum loss rate via the
CB disc can be written as:
\begin{equation}
\dot{J}_{\rm CB}=\eta\dot{M_{\rm d}}j_{\rm CB},
\end{equation}
where $\eta=\delta(t/t_{\rm vi})^{1/3}$, and $ j_{\rm CB}=\gamma
J/\mu$ is the specific orbital angular momentum of the disc
material at $r_{\rm i}$ \citep[e.g.][]{sobe97}.

The orbital evolution of BHXBs is governed by the change of the
orbital angular momentum $J$ of the system caused by the above
mentioned three mechanisms. Differentiating Eq.~(8) we get
\begin{equation}
\frac{\dot{J}}{J}=-\frac{\dot{\Omega}}{3\Omega}+\frac{\dot{M_{\rm
d}}}{M_{\rm d}}\left[1-(1-f-\delta)\frac{M_{\rm d}}{M_{\rm
BH}}-(f+\delta)\frac{M_{\rm d}}{3M}\right],
\end{equation}
where we have neglected the terms of $f\delta$. Neglect the angular
momentum loss due to gravitational wave radiation, the orbital
evolution is governed by
\begin{equation}
\frac{\dot{J}_{\rm CB}}{J} =
-\frac{\dot{\Omega}}{3\Omega}+\frac{\dot{M_{\rm d}}}{M_{\rm
d}}\left[1-\frac{M_{\rm d}}{M_{\rm BH}} + \frac{2(f+\delta)M_{\rm
d}}{3M}\right].
\end{equation}
Insert Eq.~(11) into Eq.~(13), orbit shrinking, i.e.
$\dot{\Omega}>0$, will occur when
\begin{equation}
\eta  > \frac{M_{\rm BH}}{\gamma M}\left[1-\frac{M_{\rm d}}{M_{\rm
BH}}+\frac{2(f+\delta)M_{\rm d}}{3M}\right].
\end{equation}
In this paper we take $r_{\rm i}/a=\gamma^{2}=1.7$
\citep[see][]{arty94,muno06}, then $\eta> 0.77 $ when $M_{\rm
d}\ll M_{\rm BH}$, and $f,\delta\ll 1$ (see Fig. 1).

In Fig.~1 we plot the expected relation between $\eta$ and $M_{\rm
d}$  when the orbital period is constant for BH masses $M_{\rm
BH}=7M_{\odot},10M_{\odot}$, and $20M_{\odot}$, respectively.
Generally a larger $\eta$ is required for smaller $M_{\rm d}$ and
larger $M_{\rm BH}$. This can be satisfied with an adequate mass
input rate $\delta$ and sufficiently long time $t$ of RLOF.

\begin{figure}
\begin{centering}
\epsfig{figure=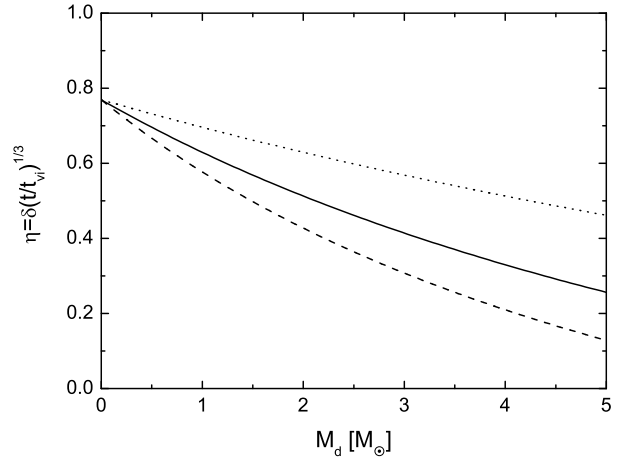, width=8cm} \caption{\label{EqmMdot} The
critical $\eta=\delta(t/t_{\rm vi})^{1/3}$ as a function of the
mass $M_{\rm d}$ of donor star for a constant binary orbit. The BH
masses are taken to be 7 $M_{\odot}$ (dashed curve), 10
$M_{\odot}$ (solid curve), and 20 $M_{\odot}$ (dotted curve),
respectively. }
\end{centering}
\end{figure}

\section[]{Numerical Results}
We have calculated the evolution of BHIMXBs adopting an updated
version of the stellar evolution code developed by
\citet{eggl71,eggl72} \citep[see also][]{han94,pols95}. In the
calculations we set initial solar chemical compositions ($X=0.7$,
$Y=0.28$, and $Z=0.02$) for the donor stars, and take the ratio of
the mixing length to the pressure scale height to be 2.0. We
include the afore-mentioned three types of orbital angular
momentum loss mechanisms during the binary evolution. For the CB
disc, we take $\alpha=0.01$ and $\beta=0.03$ \citep{bell04}.
Setting the initial mass of the BH $M^{0}_{\rm BH}=10M_{\odot}$,
we have performed the evolution calculations of BHXBs with an
initial secondary of $M_{\rm d}=3$, 4, and 5 $M_{\odot}$.

\begin{figure*}
\begin{centering}
\epsfig{figure=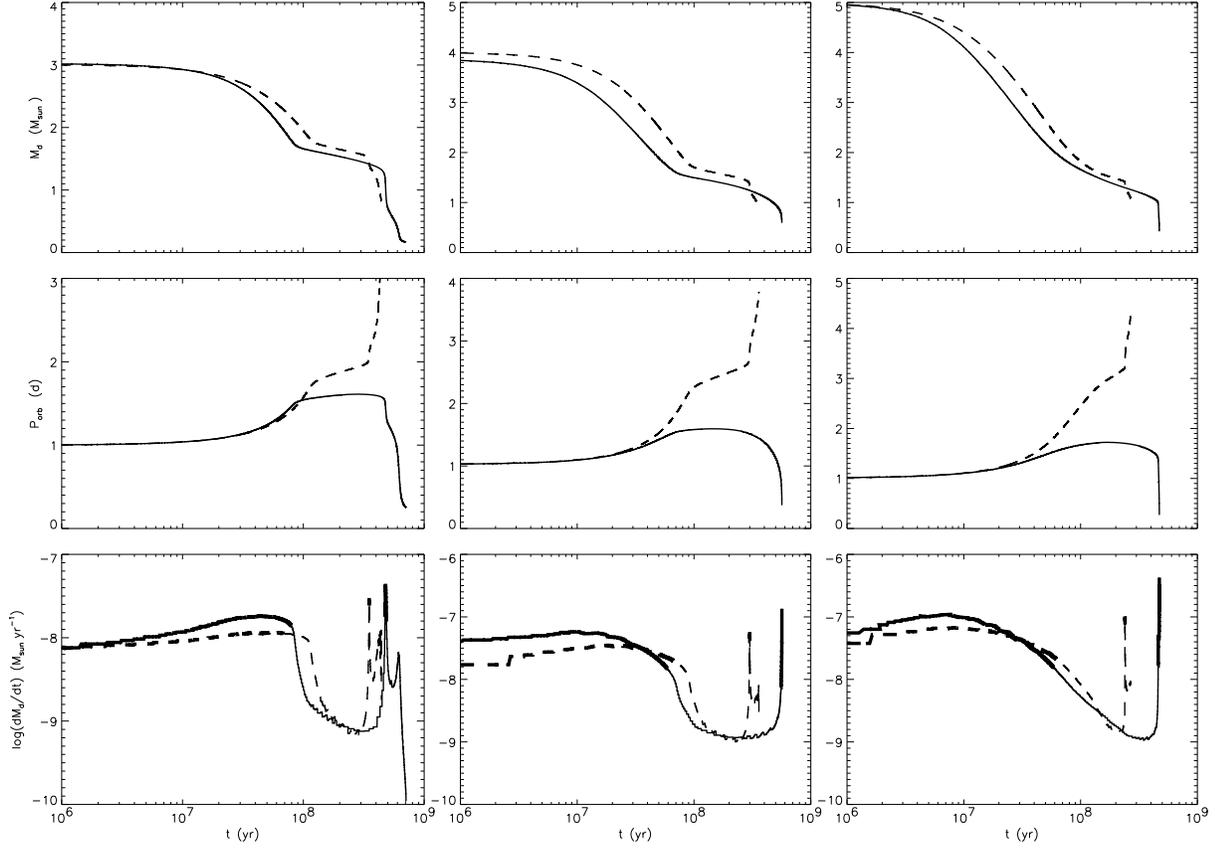, width=16cm} \caption{\label{1E6} Evolution
of the donor masses $M_{\rm d}$ (top), orbital periods $P_{\rm
orb}$ (middle), and mass transfer rates $\dot{M}_{\rm d}$ (bottom)
for BHIMXBs with the initial orbital period $P_{\rm orb}=1$ d. The
left, middle, and right panels correspond to $M_{\rm
d}=3M_{\odot}$ (solid curve $\delta=0.0055$, dashed curve
$\delta=0.004$), $4M_{\odot}$ (solid curve $\delta=0.007$, dashed
curve $\delta=0.005$), and $5M_{\odot}$ (solid curve
$\delta=0.007$, dashed curve $\delta=0.005$), respectively. Stable
and unstable mass transfer processes are plotted with thick and
thin curves, respectively.}
\end{centering}
\end{figure*}

\begin{figure*}
\begin{centering}
\epsfig{figure=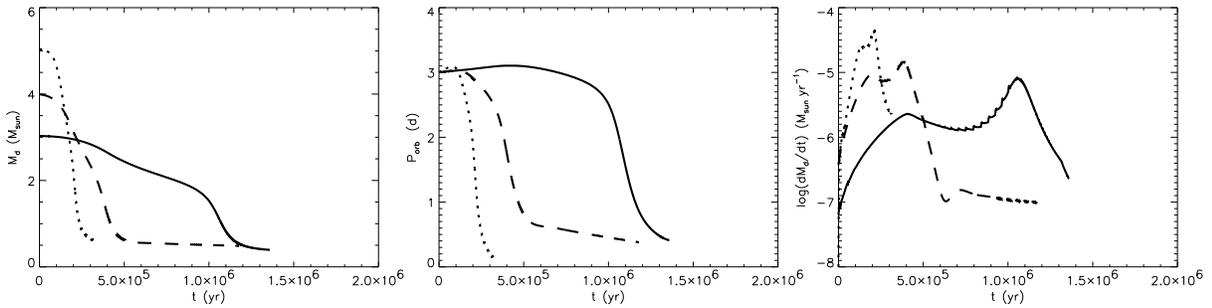, width=16cm} \caption{\label{1E6} Evolution
of the donor masses $M_{\rm d}$ (left), orbital periods $P_{\rm
orb}$ (middle), and mass transfer rates $\dot{M}_{\rm d}$ (right)
for BHIMXBs with an initial orbital period $P_{\rm orb}=3$ d. The
solid, dashed, and dotted curves correspond to the initial donor
mass of $3M_{\odot}$ ($\delta=0.055$), $4M_{\odot}$
($\delta=0.07$), and $5M_{\odot}$ ($\delta=0.09$), respectively.}
\end{centering}
\end{figure*}

We first estimate the possible range of the fractional mass input
rate into the CB disc, $\delta$, when the CB disc's influence on
the orbital  evolution of BHIMXBs is effective. From Eq.~(10) we
find the viscous timescale $t_{\rm vi}\sim 100$ yr at the inner
edge of the CB disc for a BH binary with an orbital period of
$\sim 1$ d, and the duration of RLOF $t\sim 10^{8}$ yr, so we get
$\delta\ga 0.77(t_{\rm vi}/t)^{1/3} \sim 0.008$ from Eq.~(14). For
systems with an initial orbital period of $\sim 3$ d, $t\sim
10^{6}$ yr, and $\delta \ga 0.04 $. Obviously the longer $t$, the
smaller $\delta$ required. These values of $\delta$  seem to be in
agreement with \citet{taam01} who have shown that the CB disc is
effective in draining orbital angular momentum from the system
provided that $\delta$ exceeds about 0.01. In our calculations we
find that CB discs with $\delta> \sim 0.01-0.1$ (depending on the
initial orbital periods) will cause runaway mass transfer with
high mass transfer rates, while the evolutions with $\delta$
significantly less than 0.01 is similar to those without CB discs,
as already noticed in CV evolution \citep{taam01}.

\begin{figure}
\begin{centering}
\epsfig{figure=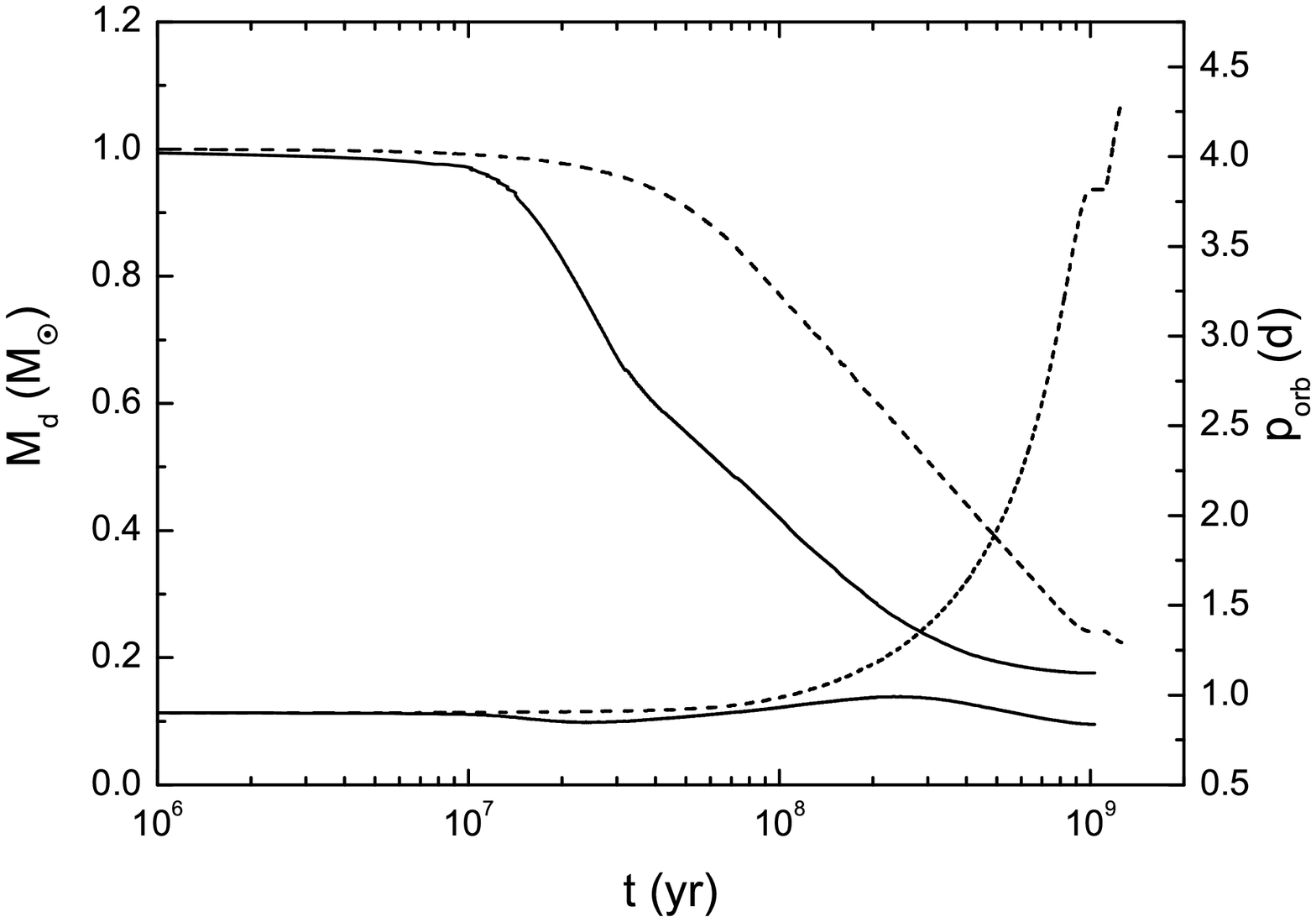, width=8cm} \caption{\label{EqmMdot} The
 evolution of the  donor masses $M_{\rm d}$ and orbital periods $P_{\rm orb}$
for NSLMXBs with a initial donor mass of $M_{\rm d}=1M_{\odot}$,
and an initial orbital period of $P_{\rm orb}=1$ d. The solid and
dashed curves correspond to $\delta=0.005$ and 0.002,
respectively. }
\end{centering}
\end{figure}

Figure 2 shows the examples of the evolutionary sequence with an
initial orbital period of 1 d. The solid and dotted lines in the
figure correspond to two different values of $\delta=(5.5-7)\times
10^{-3}$ and $\delta=(4-5)\times 10^{-3}$ respectively. As can seen
in the figure, mass transfer is initially driven by nuclear
evolution of the secondary, leading to expansion of the orbits. But
this tendency is held up when the angular momentum loss via the CB
disc becomes sufficiently strong. The mass transfer drops into a
``plateau" phase at a rate $\sim 10^{-9}M_{\odot}\rm yr^{-1}$ for a
few $10^8$ yr. These features are different from those in
\citet{pods03} for the standard evolution of BHIMXBs, in which the
orbits always increase secularly. After that the mass transfer rates
increase sharply as the secondary ascends the giant branch, but the
final orbital evolution depends on the adopted value of $\delta$.
With the higher value of $\delta$, a compact BHLMXB will be finally
produced after a few $10^8$ yr mass transfer.
If matter does not leave the CB disc, we can estimate from Fig.~2
the mass of CB disc $M_{\rm CB}\sim 0.015$, 0.028, $0.035M_{\odot}$
with the initial donor mass of 3, 4, $5M_{\odot}$, respectively.

The calculated results for slightly wider initial orbital period of
3 d are presented in Fig.~3. Generally the mass transfer rates are
higher for higher donor masses and wider initial orbits - the peaks
of the mass transfer rates can reach $\ga 10^{-5} M_{\odot}\rm
yr^{-1}$. Within less than 1 Myr the donor masses decrease to be
$\la 1 M_{\odot}$. Short orbital periods can also be attained, but
generally a larger $\delta$ (up to $\la 0.1$) is required compared
with those in Fig.~2. In this case the the CB disc is also more
massive, $M_{\rm CB}\sim 0.15$, 0.28, $0.45M_{\odot}$ with the
initial donor mass of 3, 4, $5M_{\odot}$, respectively.

One may expect that the CB disc may play a similar role in the
evolution of NSLMXBs. In Fig.~4, we plot the evolutionary sequence
of NSLMXBs with a donor star of $M_{\rm d}=1M_{\odot}$ and an
initial orbital period of $P_{\rm orb}=1$ d, to show the similar
influence of changing $\delta$ on the final orbital evolution as in
BH binaries. In our previous work on the formation of the binary
radio pulsar PSR J1713$+$0747 \citep{chen06a} we have found that for
a given final orbital period, including the CB disc can allow a more
evolved donor star at the onset of mass transfer, so that the X-ray
lifetime can be considerably decreased.

\begin{figure}
\begin{centering}
\epsfig{figure=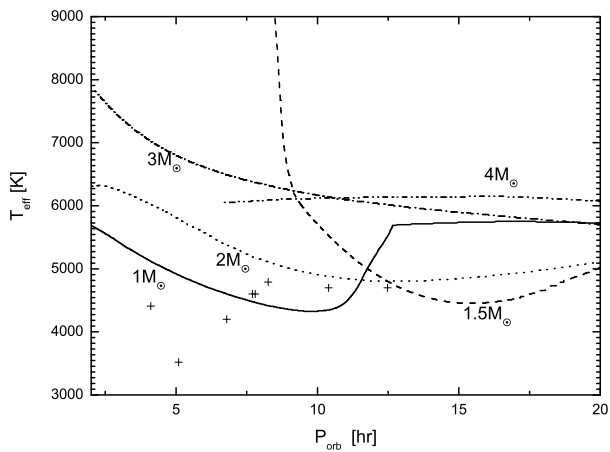, width=8cm} \caption{\label{EqmMdot} The
evolutionary tracks of BHIMXBs in the effective temperature of the
secondary vs. the orbital period plane when the initial orbital
period $P_{\rm orb,i}=1$ d. The solid, dashed, dotted, dash-dotted,
and dash-dot-dotted curves correspond to the initial donor star
masses of $1M_{\odot}$ ($\delta=0.002$), $1.5M_{\odot}$
($\delta=0.0042$), $2M_{\odot}$ ($\delta=0.0048$), $3M_{\odot}$
($\delta=0.0055$), and $4M_{\odot}$ ($\delta=0.007$), respectively.
The orbital periods of eight compact BHXBs and the earliest spectral
types of the optical companions \citep{casa06} are marked with
crosses.}
\end{centering}
\end{figure}

All of the BHLMXBs with an orbital period of $\la0.5$ d should be
transient sources \citep{casa06}. It is currently believed that
the transient behavior in X-ray binaries is due to the
thermal-viscous instability of accretion discs, when the mass
transfer rates lie below a critical value $\dot{M}_{\rm cr}$ so
that the surface temperature at the outer edge of accretion disc
is lower than the hydrogen ionization temperature \citep{king96}.
This critical mass transfer rate depends primarily on $M_{\rm
BH}$, $M_{\rm d}$, and $P_{\rm orb}$, and is given by
\citep{para96,dubu99}:
\begin{equation}
\dot{M}_{\rm cr}\simeq 8.6\times 10^{-9}\left(\frac{M_{\rm
BH}}{10M_{\odot}}\right)^{0.5}\left(\frac{M_{\rm
d}}{1M_{\odot}}\right)^{-0.2}\left(\frac{P_{\rm orb}}{1 \rm
d}\right)^{0.5} M_{\odot}\rm yr^{-1}
\end{equation}
In Fig.~2 the thick and thin curves correspond to stable and
unstable mass transfer during the evolution of BHXBs. However, the
mass transfer would be always stable for the wide orbit system
with an initial orbital period of 3 d (see Fig.~3). It is
interesting to see that the model BHLMXBs are likely to be
persistent X-ray sources. This is not compatible with the
observations of Galactic BHLMXBs, suggesting that $\delta$ does
not need to be constant but changes during the evolution (see
discussion below)\footnote{{\em Chandra} observations of the 12.6
hr ultraluminous X-ray source in the elliptical galaxy NGC3379
suggest that the current on phase has lasted $\sim 10$ yr
\citep{fabb06}. This source could be a soft X-ray transient with
long duration of outbursts. However, a persistent X-ray source as
suggested by our calculations is an alternative possibility.}.

\section{Summary and Discussion}
In this paper we have examined the influence of a CB disc on the
evolution of BHIMXBs. Assuming that a fraction $\delta$ of the
transferred material from the secondary forms a CB disc surrounding
the binary system and extracts the orbital angular momentum from the
binary, we have performed evolution calculations for BHXBs
containing a donor star of mass in the range from 3 to 5 $M_{\odot}$
with an initial orbital period of 1 d and 3 d. The calculations show
that the evolution of BHIMXB is sensitive to $\delta$. Generally a
larger $\delta$ is required for a wider initial orbit since the mass
transfer itself during the evolution always causes orbital
expansion. The results indicate that the orbits of BHXBs would show
secular shrinkage when the values of $\delta$ lie in the range of a
few $10^{-3}- 10^{-1}$. Thus our CB disc scenario suggests a new
evolutionary channel for the formation of BHLMXBs from BHIMXBs.

However, our results encounter the similar difficulty as in
\citet{just06}, that the calculated effective temperatures $T_{\rm
eff}$ are not consistent with those of the observed donor stars in
BHLMXBs. We compare the calculated results of both BH-LMXBs and
IMXBs with the observations in the $T_{\rm eff}$ - $P_{\rm orb}$
diagram in Fig.~5. One can see that the observed results seem to
be more consistent with the evolutionary tracks of original
BHLMXBs. \citet{just06} have already pointed out that this
tempreature discrepancy seems to be a generic difficulty with any
formation scenario that invokes primordially intermediate-mass
donor stars. If this effective temperature problem can be solved,
the CB disc mechanism may provide a plausible solution to the
BHLMXB formation problem, without requiring anomalous magnetic
fields in the donor stars.

The mechanism feeding the CB disc is still unclear. It has been
argued that during mass exchange in binary systems, some of the lost
matter which possesses high orbital angular momentum may form a disc
surrounding the binary system rather than leave the binary system
\citep{heu94}. This part of matter may come from the stellar wind
from the donor star, wind and/or outflow from the accretion disc, or
mass lost from the outer Lagrangian point. The values of $\delta$
adopted here seem to rule out stellar wind as the origin of the CB
disc, which requires a unreasonably high wind mass loss rate in
intermediate- and low-mass stars. We speculate that the disc
wind/outflow may play a more important role in feeding the CB disc,
as the X-ray irradiation on the accretion disc in BHXBs may
evaporate the disc much more efficiently than in CVs. This may
explain why the values of $\delta$ are about $2-3$ orders of
magnitude larger than those for CV evolution \citep{spru01}.

\citet{dubu02} investigated the structure and evolution of a
geometrically thin CB disc to calculate its spectral energy
distributions, and discussed the prospects for the detection of
such discs in the infrared and submillimeter wavelength regions.
\citet{dubu04} searched for excess mid-infrared emission due to CB
disc material in CVs. But direct detection of the CB discs in CVs
by infrared continuum studies has so far been elusive, partly
because of the lack of accurate disc atmosphere models. Recently,
\citet{muno06} studied the blackbody spectrum of BHLMXBs A 0620-00
and XTE J1118+480, and found that the inferred excess mid-infrared
emitting areas are $\sim 2$ times larger than the binary orbital
separations. Therefore, the detection of excess mid-infrared
emission from these BHLMXBs provides evidence of the existence of
CB disc around some BHLMXBs. These observations may set useful
constraints on the evolution of the CB discs. The masses of the
model CB disc are significantly larger than those ($\sim
10^{-9}M_{\odot}$) estimated by \citet{muno06}. Combined with the
non-detection of discs in CVs by \citet{dubu04}, this fact
suggests that the presence of the CB disc may not always accompany
the RLOF processes. For example, (X-ray) nova bursts may destroy
the CB disc before it has developed to be massive enough. Or
perhaps $\delta$ only needs to be high for a short time, while
normal magnetic braking takes over the CB disc during the majority
of the evolution. The latter may partly account for the
discrepancy between the lifetimes of BHLMXBs in Fig.~3 ($\sim
0.5-1.5\times 10^{6}$ yr) and those expected for observed BHLMXBs
($\sim 10^{8}-10^9$ yr). Since the CB disc can promote mass
transfer very efficiently, BHLMXBs are more likely to be observed
when the value of $\delta$ becomes extremely small.

\section*{Acknowledgments}
We thank the anonymous referee for his/her helpful comments that
significantly improved the manuscript. This work was supported by
the National Science Foundation of China (NSFC) under grant
10573010.

\bsp

\label{lastpage}

\end{document}